\documentclass{SciPost}

\hypersetup{
    colorlinks,
    linkcolor={red!50!black},
    citecolor={blue!50!black},
    urlcolor={blue!80!black}
}

\usepackage{amsmath}
\usepackage{amssymb}
\usepackage{graphicx}
\usepackage{float}
\usepackage{physics}
\usepackage{comment} %Allows to comment large sections
\usepackage{subcaption}
\usepackage{xcolor}
\newcommand{\figwidth}{0.48\textwidth}

\usepackage[bitstream-charter]{mathdesign}
\DeclareSymbolFont{usualmathcal}{OMS}{cmsy}{m}{n}
\DeclareSymbolFontAlphabet{\mathcal}{usualmathcal}

\fancypagestyle{SPstyle}{
\fancyhf{}
\lhead{\colorbox{scipostblue}{\bf \color{white} ~SciPost Physics }}
\rhead{{\bf \color{scipostdeepblue} ~Submission }}

\fancyfoot[C]{\textbf{\thepage}}
}

\begin{document}

\pagestyle{SPstyle}

\begin{center}{\Large \textbf{\color{scipostdeepblue}{
%%%%%%%%%% TODO: Write your article's title here
Unbiased Diffusion Monte Carlo for non local operators
}}}\end{center}

\begin{center}\textbf{
% Use (full) first name (+ middle name initials) + surname format.
% Separate subsequent authors by a comma, omit comma and use "and" for the last author.
% Mark the corresponding author(s) with a superscript symbol in this order
% \star, \dagger, \ddagger, \circ, \S, \P, \parallel, ...
Carlos Rodriguez Perez\textsuperscript{1,2$\star$},
Valerio Olevano\textsuperscript{2,3},
Francesco Sottile\textsuperscript{1,2} and
Vitaly Gorelov\textsuperscript{1,2}
}\end{center}

\begin{center}
{\bf 1} LSI, CNRS, CEA/DRF/IRAMIS, École Polytechnique, Institut Polytechnique de Paris, F-91120 Palaiseau, France
\\
{\bf 2} European Theoretical Spectroscopy Facility (ETSF)
\\
{\bf 3} Institut Néel CNRS/UGA UPR2940,
25 Rue des Martyrs, 38042 Grenoble, France
%%%%%%%%%% END TODO: AFFILIATIONS
% Provide email address of corresponding author(s)
\\[\baselineskip]
$\star$ \href{mailto:carlos.rodriguez-perez@polytechnique.edu}{\small carlos.rodriguez-perez@polytechnique.edu}
\end{center}

\section*{\color{scipostdeepblue}{Abstract}}
\textbf{\boldmath{%
We propose a new mathematically exact method for computing unbiased 
Diffusion Monte Carlo (DMC) estimates of non-local operators. 
We demonstrate that the current state-of-the-art technique, 
Forward Walking, is only exact for local quantities and fails to 
yield unbiased results for the non-local components of reduced density 
matrices (RDMs). Our method significantly outperforms Forward Walking, as shown in two systems: in the symmetric Hubbard dimer it yields a pure 1RDM; while in the Helium atom it will give an unbiased 1RDM in the limits of zero time step and infinite walkers.
}}

\vspace{\baselineskip}

%%%%%%%%%% BLOCK: Copyright information
% This block will be filled during the proof stage, and finilized just before publication.
% It exists here only as a placeholder, and should not be modified by authors.
\noindent\textcolor{white!90!black}{%
\fbox{\parbox{0.975\linewidth}{%
\textcolor{white!40!black}{\begin{tabular}{lr}%
  \begin{minipage}{0.6\textwidth}%
    {\small Copyright attribution to authors. \newline
    This work is a submission to SciPost Physics. \newline
    License information to appear upon publication. \newline
    Publication information to appear upon publication.}
  \end{minipage} & \begin{minipage}{0.4\textwidth}
    {\small Received Date \newline Accepted Date \newline Published Date}%
  \end{minipage}
\end{tabular}}
}}
}
%%%%%%%%%% BLOCK: Copyright information

% For convenience during refereeing we turn on line numbers:
%\linenumbers

%\vspace{10pt}
%\noindent\rule{\textwidth}{1pt}
%\tableofcontents
%\noindent\rule{\textwidth}{1pt}
%\vspace{10pt}

% Write your article contents here, starting from first \section.
% An example structure is given below.

\section{Introduction}
\label{sec:intro}
% TODO: write your article here.
Quantum Monte Carlo (QMC) methods have established themselves as 
some of the most accurate tools for studying many-body quantum systems 
\cite{foulkes,lucia_book}. Among these, Diffusion Monte Carlo (DMC) 
\cite{kalos1970_DMC} is widely used to obtain high-precision benchmarks 
for atoms, molecules, and solids 
\cite{yang2020electronic, filippi1999quantum, yang2020quantum, 
nemec2010benchmark, leung1999calculations}, often in excellent 
agreement with experimental data. Due to this accuracy, DMC is 
frequently employed to produce gold-standard benchmarks and fit 
density functionals \cite{Perdew_Wang_egas_correl_energy, 
sottile2001DMC_jellium, hongo2006diffusion_atom, Senatore_Ceperley_Response}.

Despite its success with total energy calculations, a lot of effort has been given to improving the quality of DMC results for other observables. These efforts have historically been focused on several fronts: mitigating time-step errors 
\cite{umrigar1993diffusion, anderson2024reducing}, refining the 
fixed-node approximation through backflow correlations 
\cite{kwon1994quantum, lopez2006inhomogeneous}, and addressing 
finite-size effects \cite{lin2001twist, chiesa2006finite, 
drummond2008finite,holzmann2016theory}. However, a more subtle 
challenge arises from Mixed Estimator (ME) effects, where DMC samples 
the distribution $\psi_T^*\psi_0$ rather than the pure $\psi_0^2$. 
While other projector methods like Reptation QMC 
\cite{baroni1999reptation} sample the pure distribution directly, 
or like Auxiliary Field QMC \cite{zhang2003quantum} that removes the
bias via back-propagation, DMC typically relies on Forward Walking (FWD) 
\cite{liu1974quantum_fwd}, correction estimates 
\cite{ceperley1979quantum}, or other techniques
\cite{zhang1993bilinear, casulleras1995unbiased},  
to recover pure estimates. One vital operator that suffers from ME effects is the one body reduced density matrix (1RDM).

The 1RDM is a fundamental quantum object that 
provides critical insights, from the nature of chemical bonds \cite{wutig}, 
to metal-insulator phase transitions  \cite{pierleoni2018electron}. 
Consequently, there is a renewed interest in computing highly accurate 1RDMs using QMC \cite{AYOUB_kohn_sham, ghaffar2026many, 
wines2020first, wines2023systematic,
kaiser2025accurate, assaraf2007improved}. The 1RDM is defined as
\begin{equation}
\gamma_0^{(1)}(\mathbf{r},\mathbf{r}') = N_e \int d\mathbf{r}_2...d\mathbf{r}_N \psi_0^*(\mathbf{r},\mathbf{r}_2...\mathbf{r}_N)\psi_0(\mathbf{r}',\mathbf{r}_2...\mathbf{r}_N) \ ,
\label{eq_1rdm_def}
\end{equation}
where $\psi_0$ is the many-body ground state wave function and $N_e$ is 
the number of electrons in the system. 

Crucially, while FWD can make DMC statistically exact for local 
quantities such as the density, it remains insufficient for non-local operators like the 
1RDM. This limitation arises because non-local estimators require the 
explicit evaluation of the trial wave function $\psi_T$ within the 
operator's accumulator. Even with a perfect distribution of walkers, 
this "accumulator bias" persists. This issue has been previously noted 
in the context of bosonic systems \cite{moroni2004condensate, 
buonaura1998numerical}, yet a general solution for electronic systems 
has remained elusive. While other Projector Monte Carlo techniques like AFQMC or RQMC can yield unbiased 1RDMs \cite{zhang2003quantum,carleo2010reptation}, there is no scheme within DMC that treats this accumulator bias.

In this work, we propose a new technique based on an extension of the FWD 
principle that is capable of yielding pure estimators for RDMs within DMC by removing 
bias from both the distribution and the accumulator. We benchmark our 
approach using the helium atom and the symmetric Hubbard dimer, 
demonstrating that our method recovers the exact 1RDM in the latter and 
significantly improves upon FWD in the former. The 1RDM observable reveals an error related to the slow convergence of the distribution with time step. 

The article is organized as follows: we first detail the methods for treating 
ME and accumulator bias, then present the results for the helium atom and the 
Hubbard dimer, and finally discuss the implications and extensions of our 
technique.

\section{Methods}
Quantum Monte Carlo methods compute expectation values of operators by stochastically solving the Schrödinger equation. The simplest approach is Variational Monte Carlo (VMC), where the Variational Theorem is employed to optimize an explicitly correlated electronic wave function. Then this trial wave function $\psi_T$ is employed for the evaluation of other operators at the variational level
\begin{equation}
    \mathcal{O}_{T} = \frac{\bra{\psi_T} O \ket{\psi_T}}{\bra{\psi_T} \ket{\psi_T}}\ .
    \label{equation_Ot_definition}
\end{equation}
VMC is inherently limited by the quality and flexibility of our proposed trial wave function. This bias on the choice of $\psi_T$ can be partially removed by using Diffusion Monte Carlo (DMC).

DMC solves the imaginary time Schrödinger equation by the repeated application of an approximate Green's function \cite{kalos1970_DMC, anderson1975random}. This procedure evolves the distribution of walkers from the variational $\psi_T^*\psi_T$ to the DMC mixed distribution $\psi_T^*\psi_0$. Computing operators within DMC yields the mixed estimate
\begin{equation}
     \mathcal{O}_{M} = \frac{\bra{\psi_T} O \ket{\psi_0}}{\bra{\psi_T} \ket{\psi_0}}\ .
    \label{eq_Ov_Om}
\end{equation}
Here $\psi_0$ is the best possible approximation to the real ground state wave function which has the same nodal structure as the proposed $\psi_T$. This is called fixed node approximation and it is a source of error in DMC simulations. In this work we focus on ME bias, so we chose examples where the wave function has no nodes. 

Objects that commute with the Hamiltonian are not affected by this mixed character, so they yield exact estimates $\mathcal{O}_0$. Quantities that do not commute with the Hamiltonian give the mixed estimator $\mathcal{O}_{M}$, which still depends on the quality of $\psi_T$. Forward walking and its approximate correction estimates are able to correct the mixed estimator error.

\begin{figure}[H]
  \begin{center}
  
  \includegraphics[width=0.94\textwidth]{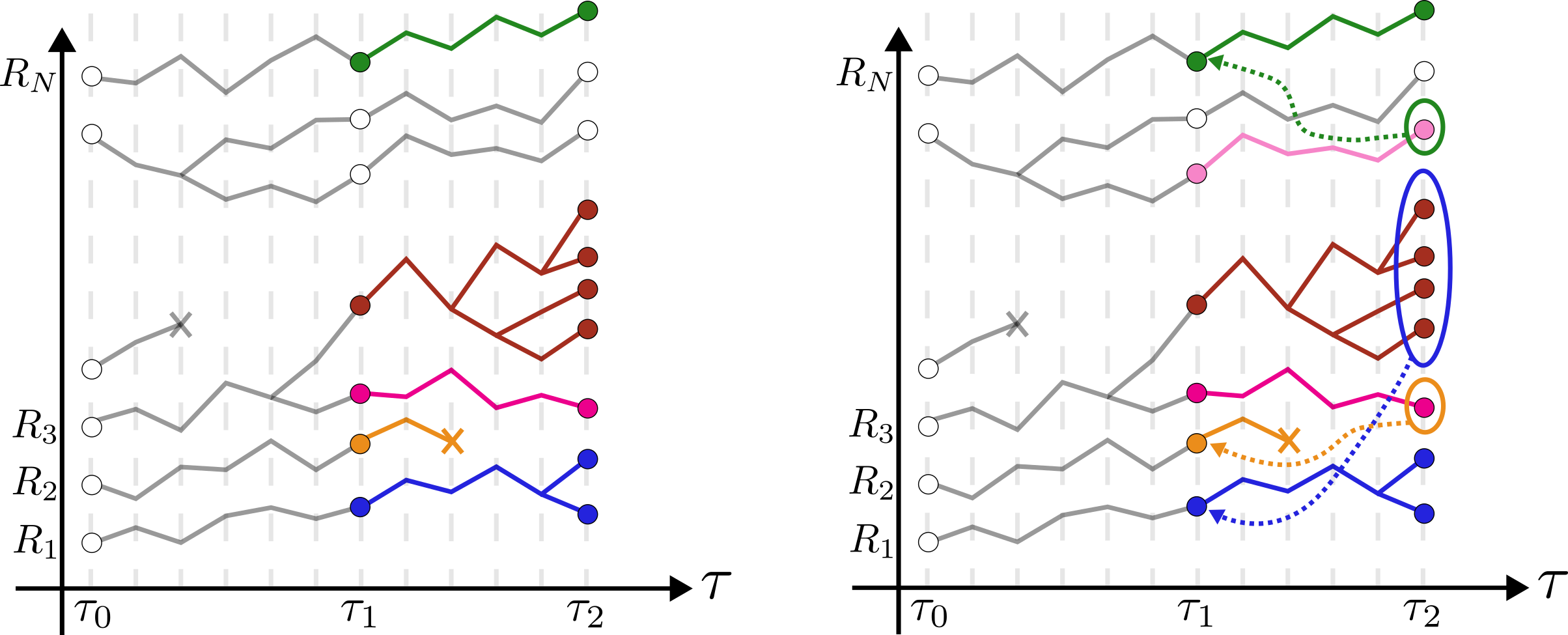}
  \end{center}
	\caption{Schematic representation of FWD (left) and AUX (right). Left figure represents forward walking, where after an equilibration time $\tau_1$ we can perform DMC, or project to a time $\tau_2=\tau_1+\tau_{fwd}$ and use the descendants $d(\mathbf{R}_i)$ of each walker $\mathbf{R}_i$ as a weight (see Eq.\eqref{eq_fwd_definition}). For example, the blue walker has a weight of $2$, the red has $4$, etc. On the right we have Auxiliary walking, which reweighs each walker $\mathbf{R}_i$ according to the descendants of the auxiliary walker $\mathbf{R}_i'$ (see Eq.\eqref{eq_aux_definition}). We assume that we found $\mathbf{R}_1' = \mathbf{R}_4$, with this we can reweigh walker $\mathbf{R}_1$ by $d(\mathbf{R}_1')$, which happens to be $d(\mathbf{R}_4)$. This means that the blue walker now has a weight of $4$ instead of $2$. 
    }
  \label{Figure_schematic_AUX}
\end{figure}

Correction estimates are simple formulas that partially remove the error introduced by the mixed distribution \cite{ceperley1979quantum}. One may use the Subtraction formula (SUB) or the Division formula (DIV), both derived in Appendix \ref{App_correction_estimates}. 

\begin{equation}
	\mathcal{O}_{\text{\tiny SUB}} \sim  2\mathcal{O}_{M}-\mathcal{O}_{T} \quad \mathrm{and}\quad \mathcal{O}_{\text{\tiny DIV}} \sim  \mathcal{O}_{M}^2/\mathcal{O}_{T}\ .
    \label{eq_correction_estimates}
\end{equation}
Forward walking (FWD) is a technique first proposed by Liu, Kalos and Chester in 1974 which fully 
removes the mixed distribution bias \cite{liu1974quantum_fwd}. 
It achieves this by reweighting the mixed estimate contribution of each walker $\mathbf{R}_i$ according 
to the future number of descendants $d(\mathbf{R}_i)$ of that particular walker 
(throughout this work weights $d$ will be assumed normalized). 
\begin{equation}
    \mathcal{O}_{F} = \sum_i d(\mathbf{R}_i) \left(\frac{\bra{\psi_T} O  \ket{\psi_0}}{\bra{\psi_T} \ket{\psi_0}}\right)_i = \sum_i d(\mathbf{R}_i) \mathcal{O}_{M}(\mathbf{R}_i).
    \label{eq_fwd_definition}
\end{equation}
This is illustrated in the left panel of Fig.~\ref{Figure_schematic_AUX}.
The number of descendants is obtained through the branching algorithm. This procedure exploits the fact 
that the asymptotic number of descendants $d(R_i)$ of a walker $\mathbf{R}_i$ is proportional to the ratio 
$\psi_0(\mathbf{R}_i)/\psi_T(\mathbf{R}_i)$ \cite{liu1974quantum_fwd, lucia_book}. The reweighting by 
$d(\mathbf{R}_i)$ gives averages as if the walkers were distributed according to $\psi_0^2$, 
removing all $\psi_T$ bias in the distribution. 

While the correction estimates (both DIV and SUB) are only first order approximations, the FWD approach 
is effective in removing the ME error. However, this paper demonstrates that the latter still suffers 
from a new source of error, intimately related to the non-local character of the observable, that contains 
an explicit dependance on the $\psi_T$, so biasing the result. 
We discuss this now by taking the most important example, the 1RDM. 

The 1RDM can be computed by QMC methods by rewriting Eq.\eqref{eq_1rdm_def}, as firstly done by W. L. 
McMillan \cite{mcmillan1965ground}.  Within QMC the ground state 1RDM is 
\begin{equation}
   \gamma^{(1)}_0(\mathbf{r},\mathbf{r}') =N_e \expval{\frac{\psi_0(\mathbf{r}',...)}{\psi_0(\mathbf{r}_1,...)}\delta(\mathbf{r}-\mathbf{r}_1)}_{\psi_0^2} \ .
   \label{eq_1rdm_gs}
\end{equation}
Which we must approximate within QMC via
\begin{equation}
   \gamma^{(1)}_{QMC}(\mathbf{r},\mathbf{r}') =N_e \expval{\frac{\psi_T(\mathbf{r}',...)}{\psi_T(\mathbf{r}_1,...)}\delta(\mathbf{r}-\mathbf{r}_1)}_{\Pi} \ ,
   \label{eq_1rdm_MIXED}
\end{equation}
where the distribution $\Pi$ changes for each method: VMC ($\Pi=\psi_T^*\psi_T$), DMC ($\Pi=\psi_T^*\psi_0$) 
or FWD ($\Pi=\psi_0^*\psi_0$). It can be seen that computing the 1RDM requires the explicit evaluation of 
wave functions. In general one does not have access to the ground state wave function, so these ratios must 
be evaluated from the trial wave function $\psi_T$. Even using the best method available (FWD), our 1RDM 
(or other estimates that require the explicit evaluation of wave functions) will be biased by the choice of 
$\psi_T$. When computing non-local operators, the SUB and DIV corrections will approach the FWD estimate, 
as they only correct the distribution but not the underlying accumulator 
(See Appendix \ref{App_correction_estimates}). 

This paper proposes a new method capable of fully removing the trial wave function bias, both in the distribution and in the accumulator. Two distinct sources of error are present in the accumulator of the : the non-locality means we need to evaluate $\psi_T(\mathbf{R}')$ in the accumulator for a walker at position $\mathbf{R}$; and a factor $1/\psi_T(\mathbf{R})$.

Imagine that an auxiliary walker $\mathbf{R}_i' = \{\mathbf{r}',\mathbf{r}_2...\mathbf{r}_N \}$ could be found (or created) for each walker $\mathbf{R}_i = \{\mathbf{r}_1,\mathbf{r}_2...\mathbf{r}_N \}$ so that they differ only in the position of one electron. Our proposal, Auxiliary Walking (AUX), exploits the fact that the number of descendants $d(\mathbf{R}'_i)$ at position $\mathbf{R}'_i$ are proportional to $\psi_0(\mathbf{R}'_i)/\psi_T(\mathbf{R}'_i)$. This is exactly the weight we need to introduce to remove the non-local part of the bias (the numerator) of the DMC estimate in Eq.\eqref{eq_1rdm_MIXED}. An illustration of this technique is shown in the right pannel of Fig.~\ref{Figure_schematic_AUX}.

The weight we have to introduce to cure the other $\psi_T$ bias (the denominator coming from the 1RDM accumulator structure) is $[d(\mathbf{R}_i)]^{-1}$, which is the inverse of the one needed to correct the mixed distribution, $d(\mathbf{R}_i)$. There is a cancellation of error between the $\psi_T$ in the denominator and in the distribution. The introduction of the auxiliary weight $d(\mathbf{R}'_i)$ is enough to yield unbiased 1RDMs. This results in an equation similar to that of FWD,
\begin{equation}
    \mathcal{O}_{A} 
    = \sum_i\ d(\mathbf{R}'_i)\ \mathcal{O}_{M}(\mathbf{R}_i) = \mathcal{O}_0 \ .
    \label{eq_aux_definition}
\end{equation}
A straightforward extension of this technique would allow to compute pure estimators for other non-local observables, like the momentum distribution or even the two body RDM. We now want to quantify the accuracy of these approaches for the ME and accumulator bias. To do this, we chose two systems with an exact solution: one is realistic, the Helium atom (for which a numerical solution can be evaluated to an arbitrary precision); and one is a model, the Hubbard dimer (which gives the double advantage of an analytic solution and it is free from 
time-step error). 

\section{Results}
The helium atom is one of the most interesting playgrounds for many-body physics. It is one of the few realistic systems with interacting electrons for which a numerical expression of its ground state wave function is known \cite{Hylleraas, SciPostPhys.6.4.040}. This gives us the possibility to benchmark QMC against exact solutions. The exact ground state is strictly positive, which means we do not have fixed node errors. This makes the helium atom a perfect system on which to test the accuracy of Diffusion Monte Carlo as the energy only suffers from the much treatable time step error.

\begin{figure}[h]
  \centering
  % First Subfigure
  \begin{subfigure}[b]{0.49\textwidth}
    \centering
    \includegraphics[width=\textwidth]{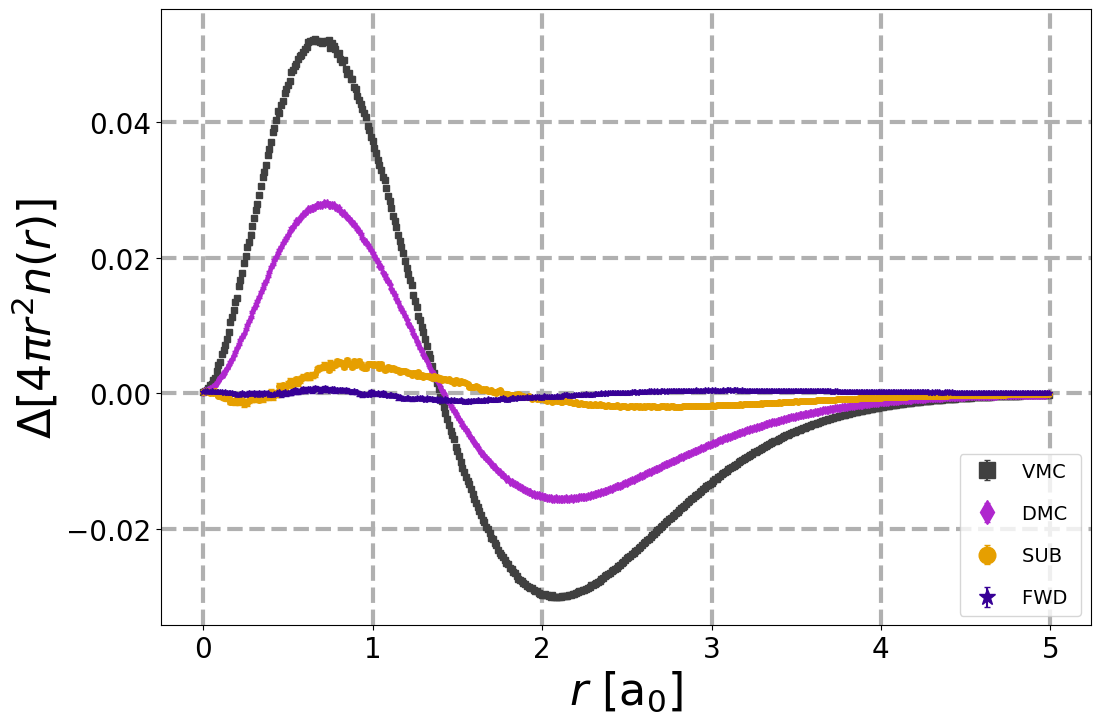}
  \end{subfigure}
  \hfill 
  \begin{subfigure}[b]{0.49\textwidth}
    \centering
    \includegraphics[width=\textwidth]{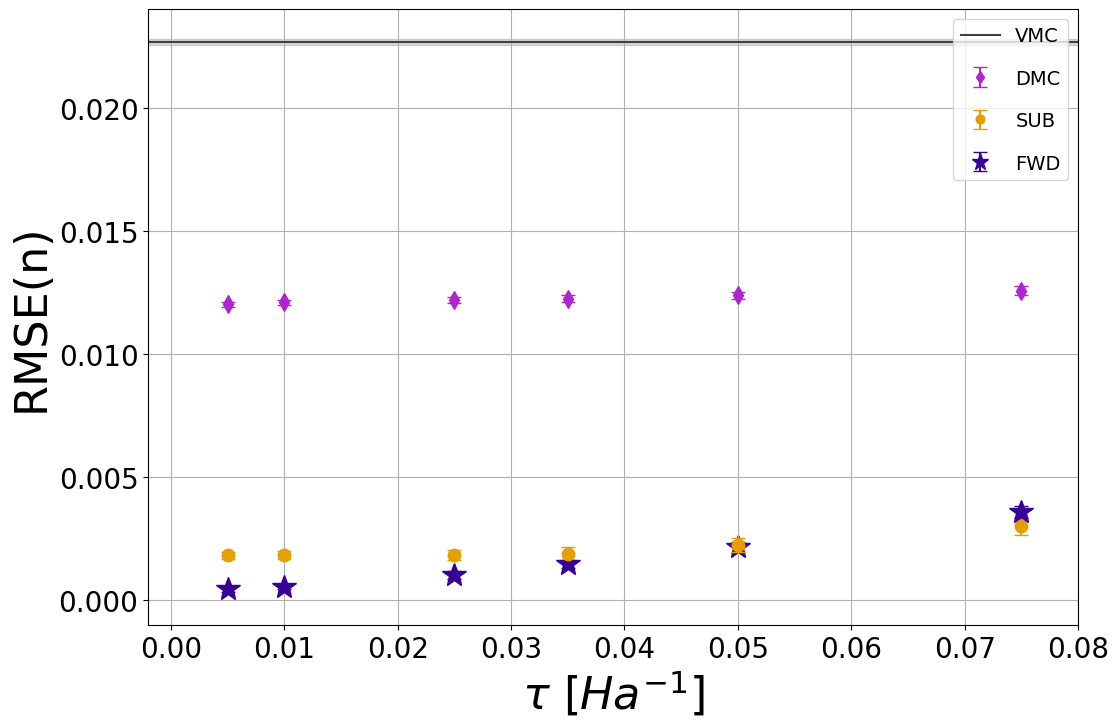}

  \end{subfigure}
  
  % Main overall caption for both figures
	\caption{QMC results for the density, we show VMC (grey), DMC (purple), SUB (orange) and FWD (blue). a$)$ Real space deviation from the exact density, DMC and FWD performed at a time step of $\tau = 0.005$. b$)$ Root Mean Square Error (RMSE) of the density against the time step.}
  \label{Figure_density}
\end{figure}

We first decided to compute the electronic density of the helium atom $n(\mathbf{r}) = \sum_i \expval{\delta(\mathbf{r}_i-\mathbf{r})}$, which due to the symmetries of the problem can be taken as spherically symmetric. In Fig.~\ref{Figure_density} we show calculations with VMC (grey), DMC (purple), SUB (orange) and FWD (blue). In the left panel we show the difference of the QMC densities with respect to the exact one, while in the right panel we show the Root Mean Square Error (RMSE) of the density for several time steps. We can see that DMC represents a large improvement compared to VMC, taking results roughtly half way to the exact density. Time step reduction barely improves DMC densities, showing that Mixed Estimator effects dominate on DMC densities. The SUB correction gives a meaningful improvement on top of DMC at little extra cost, but is also limited as it does not converge to the exact density at zero time step. The FWD density is almost exact, with a very small deviation caused by a residual time step error. In the right panel we can see that upon time step extrapolation $\tau\rightarrow 0$, FWD yields a statistically exact density. We have performed two checks on our calculations: first, DMC energies converge to the exact one as we approach $\tau \rightarrow 0$; our high quality FWD densities produce an almost exact Hartree energy.

We then moved on to compute the 1RDM. Since it is a 6 dimensional object, we decided to restrict our calculations by fixing $\mathbf{r}' = (0.5,0,0)$. Also, we only present the line $\mathbf{r}=(X,0,0)$ for visual clarity as this is sufficient to identify the non-exactness of FWD for the 1RDM. The choice of 0.5 for the second position was motivated by the large number of walkers in that region, helping make the calculation more efficient. It also helped us see slight asymmetries in the 1RDM along the $X$ axis. All 1RDMs we present have been normalized by the condition $\gamma^{(1)}(\mathbf{r}';\mathbf{r}')=n(\mathbf{r}')$.

\begin{figure}[h]
  \centering
  % First Subfigure (Fig A)
  \begin{subfigure}[b]{0.49\textwidth}
    \centering
    \includegraphics[width=\textwidth]{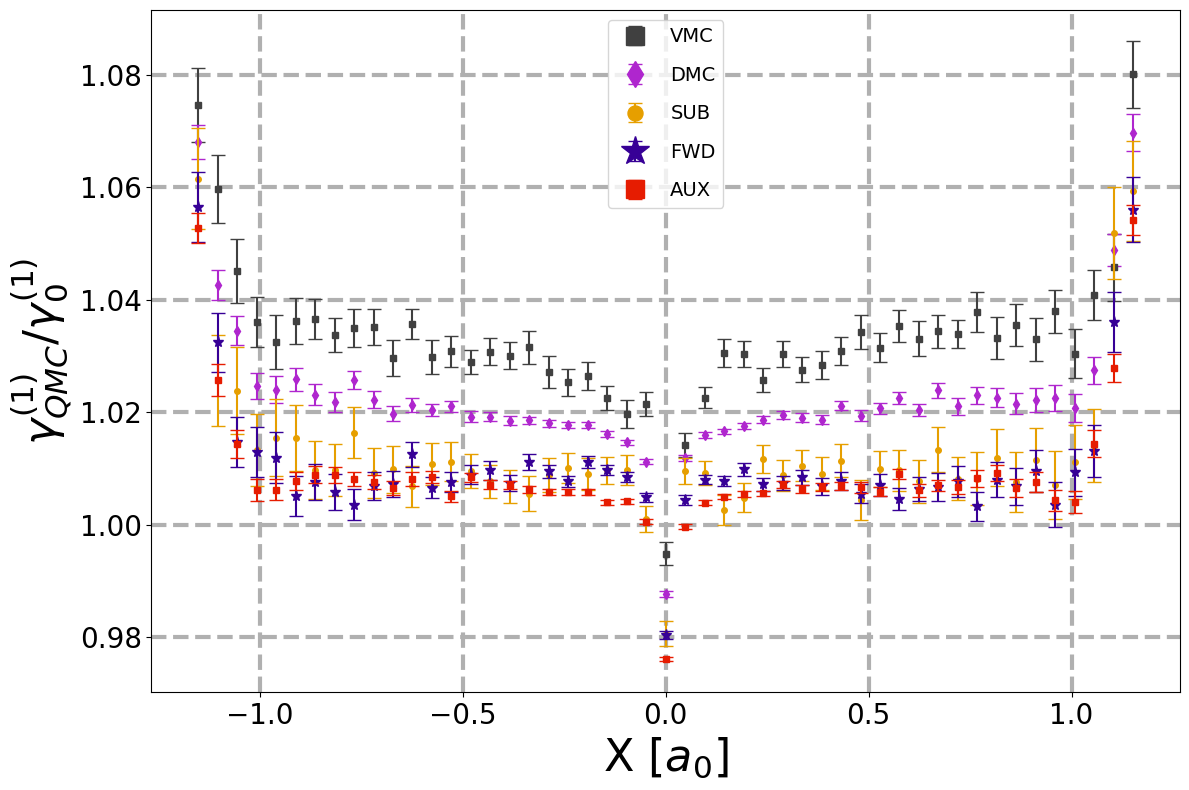}
  \end{subfigure}
  \hfill 
  \begin{subfigure}[b]{0.49\textwidth}
    \centering
    \includegraphics[width=\textwidth]{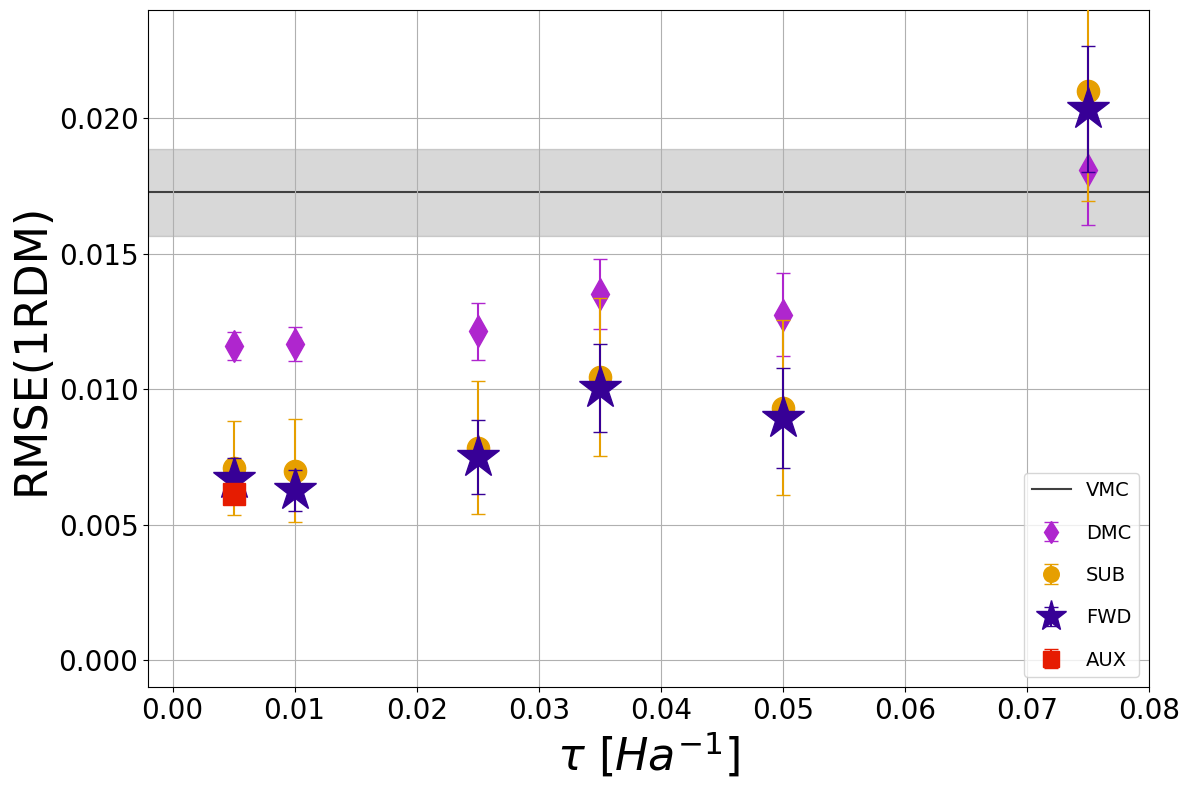}

  \end{subfigure}
  
  % Main overall caption for both figures
	\caption{QMC results for the 1RDM along the line $\gamma^{(1)}(X,0,0;0.5,0,0)$, we show VMC (grey), DMC (purple), SUB correction (orange), FWD (blue) and AUX (red). a$)$ Real space 1RDM deviation from exact result (at a time step of $\tau = 0.005$). b$)$ Root Mean Square Error (RMSE) of the 1RDM along the previous line.
  }
  \label{Figure_1rdm}
\end{figure}

In Fig.~\ref{Figure_1rdm} we illustrate our calculations of the 1RDM with VMC (grey), DMC (purple), SUB (orange), FWD (blue) and AUX (red), with the left panel showing a real space cut, and the right panel an RMSE analysis with time step. DMC improves on VMC, but not as much as for the density. This is due to the presence of bias beyond the mixed distribution, i.e. the $\psi_T$ bias in the accumulator. The SUB results suffer the same issue, they are constructed to improve the distribution, so they outperform DMC, but not as much as in the density. FWD gives slightly better results than SUB, but just marginally, the reason is again that FWD is designed to fully remove the bias in the distribution, not in the accumulator (See Eq.\eqref{eq_1rdm_MIXED}). Still, SUB and FWD give high quality 1RDMs, this stems from the fact that the distribution is corrected and that the accumulator's $\psi_T$ bias is partially cancelled between the numerator and denominator. These results suggest that the accumulator bias is of large importance if pure 1RDMs are required. We have performed only one time step calculation of AUX because at low $\tau$ is where we expect the biggest difference with respect to FWD. As illustrated in Fig.~\ref{Figure_1rdm}, AUX outperforms FWD by roughtly the same amount as the much studied improvement of FWD on top of the SUB correction, this allows AUX to give us the best 1RDM attainable within DMC. Still, AUX falls short of providing an exact 1RDM in this system. There is an extra source of error which we must investigate, as our technique is in principle exact. For this we will turn to the Hubbard dimer. 

The symmetric Hubbard dimer with two opposite spin electrons is exactly solvable \cite{el2024total} and can shed light on the nature of this bias. Since the electrons are restricted to just four configurations, finding the corresponding auxiliary walker for AUX is very simple. The Hamiltonian is
\begin{equation}
    H = \varepsilon_0 \sum_{i,\sigma} n_{i,\sigma} - t \sum_\sigma (c_{1,\sigma}^\dagger c_{2,\sigma} + c_{2,\sigma}^\dagger c_{1,\sigma}) + U\sum_i n_{i,\uparrow} n_{i,\downarrow}
    \label{eq_symm_hub_dimer}
\end{equation}
where $c^\dagger$ and $c$ are the creation and annihilation operators, $n_{i,\sigma}$ is the on site spin density ($c_{i,\sigma}^\dagger c_{i,\sigma}$), $\sigma$ is the spin index, $\varepsilon_0$ is the level energy (set to 0), $t$ the hopping parameter and $U$ the on-site interaction.

\begin{figure}[h]
  \centering
  \includegraphics[width=0.78\textwidth]{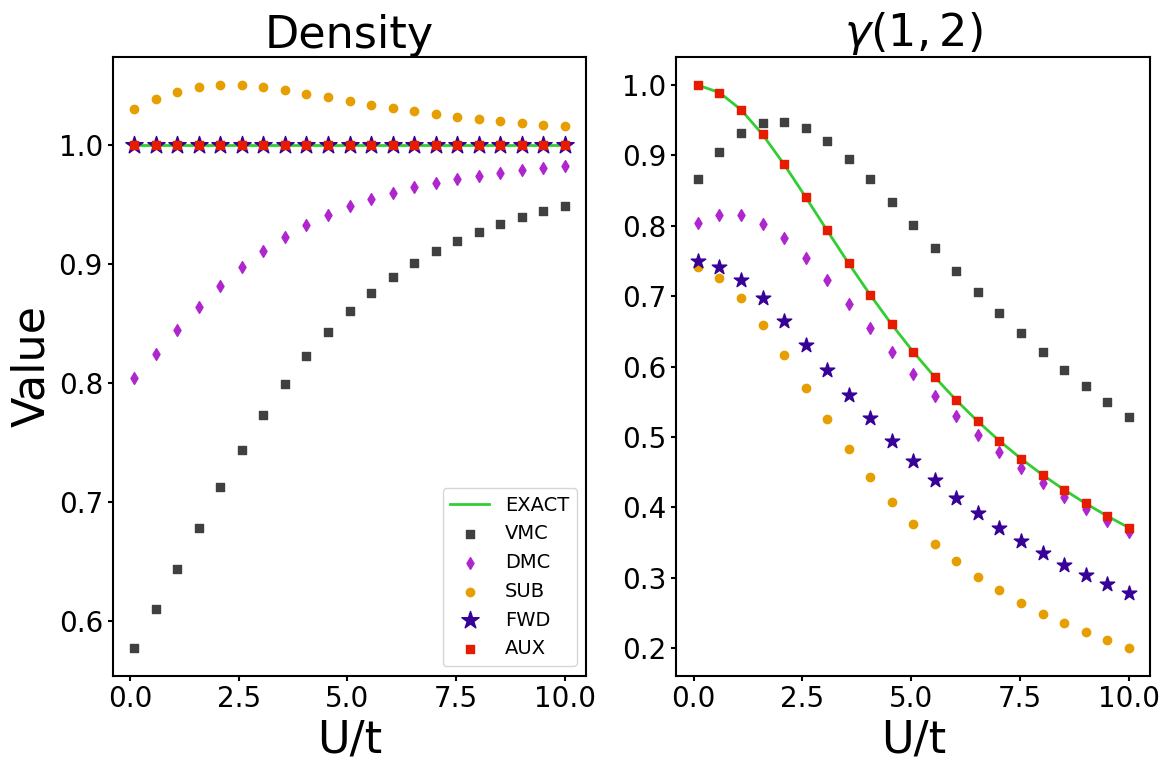}
  \caption{Density and off diagonal element of the 1RDM of the symmetric Hubbard dimer against the ground state solution (green). We show VMC (grey), DMC (purple), SUB (orange), FWD (blue) and AUX (red). Error bars are negligible.
  }
  \label{fig_toverU}
\end{figure}

In the Hubbard dimer the 1RDM is a 2x2 matrix. In Fig.~\ref{fig_toverU} we show the density (diagonal of the 1RDM) and one non-local (off diagonal) element of the 1RDM for VMC (grey), DMC (purple), SUB (orange), FWD (blue) and AUX (red). We can see that when computing the density, both FWD and AUX are exact (as AUX reduces to FWD in the local part of the 1RDM). The non-local part is different, here FWD fails to correctly describe the 1RDM, while AUX yields a pure result. Of all the methods we present, only AUX yields the exact 1RDM for all values of correlation studied.

In the Hubbard dimer we are able to perfectly sample the mixed distribution $\psi_T^*\psi_0$, without time step bias and with a number of walkers which is effectively infinite. This tells us that the error present in the AUX calculation of Fig.\ref{Figure_1rdm} lies in the distribution of walkers. AUX requires a cancellation of $\psi_T$ error between the denominator of the 1RDM and the distribution of walkers, which happens only if the walkers are distributed exactly as $\psi_T^*\psi_0$. We have performed a study of this cancellation of error in the helium atom in Appendix \ref{App_psiTcancel}. There we show that this cancellation partially takes place, but that it converges extremely slowly with time step. This makes AUX very sensitive to errors in the distribution of walkers. Still, AUX outperforms FWD in the helium atom, showing that this lack-of-cancellation remaining error is smaller than the accumulator bias in FWD. 
And, as shown in the Hubbard dimer, AUX is exact in the $\tau \rightarrow 0$ and $N_{walkers} \rightarrow \infty$ limit.

\section{Conclusion}
In this work, we have demonstrated that the standard state-of-the-art 
techniques in DMC, specifically Forward Walking and subtraction 
correction estimates, are fundamentally unable to yield exact one-body 
reduced density matrices (1RDMs). We introduced Auxiliary Walking (AUX), 
a novel reweighting technique mathematically designed to remove trial 
wave function bias from both the distribution and the accumulator. 
The exactness of AUX was rigorously proven through a proof-of-concept 
in the symmetric Hubbard dimer, where it recovered the pure 1RDM across 
all correlation regimes.

Our application of AUX to the helium atom allowed us to characterize 
a critical issue related to the time step error: the slow convergence of the $\psi_T$ 
cancellation between the operator's denominator and the walker 
distribution. We show that this cancellation is highly sensitive to 
the time step (Appendix \ref{App_psiTcancel}), 
explaining why pure estimates are more elusive in 
realistic systems than in the Hubbard dimer. However, the fact that AUX 
consistently outperforms FWD in the helium atom confirms that the 
accumulator bias is the primary bottleneck for non-local operators 
in DMC.

This work establishes a framework for obtaining unbiased non-local 
observables within the Diffusion Monte Carlo method. While the 
helium atom presents a challenging case due to its small number of 
electrons, we anticipate that the method will be more robust in 
larger systems where the relative impact of local wave function 
fluctuations is reduced. The ability to compute pure 1RDMs opens 
a path toward evaluating a wide array of 1RDM-based functionals 
and quantum properties at the highest accuracy achievable with 
projector Monte Carlo methods.

\section*{Acknowledgements}

% TODO: include author contributions
\paragraph{Author contributions}
All authors contributed equally to the article. 

% TODO: include funding information
\paragraph{Funding information}
C.R.P. acknowledges that this work has been carried out at the Energy4Climate Interdisciplinary Center (E4C) of IP Paris, which is in part supported by 3rd Programme d’Investissements d’Avenir [ANR-18-EUR-0006-02] and the ANR France 2030 ExcellenceS Programme [ANR-22-EXES-00013]. 
This project was provided with computer and storage resources by GENCI at IDRIS and TGCC thanks to the grant 2026-0544.

\begin{appendix}
\numberwithin{equation}{section}

\section{Derivation of the correction estimate}
\label{App_correction_estimates}
In this appendix we will derivate the formula for the subtraction (SUB) and division (DIV) correction to the 1RDM. We will use
\begin{equation}
    \psi_0 (R) = \psi_T(R) [1+\Delta(R)] \ ,
    \label{equation_corr_estimate_wf}
\end{equation}
where $\Delta(R)$ is assumed small. Using this we can express our mixed DMC estimate as,
\begin{align}
    \nonumber \gamma^{(1)}_M(r,r') &= N_e \expval{\frac{\psi_T(r',...)}{\psi_T(r_1,...)}\delta(r-r_1)}_{\psi_T\psi_0} \\
    \nonumber &= N_e \int dR\ \frac{\psi_T(r',...)}{\psi_T(r_1,...)}\delta(r-r_1)\psi_T(R)\psi_0(R) \\
    \nonumber &= N_e \int dR\ \frac{\psi_T(r',...)}{\psi_T(r_1,...)}\delta(r-r_1)\psi_T(R)\psi_T(R) [1+\Delta(R)] \\
    \nonumber &= N_e \expval{\frac{\psi_T(r',...)}{\psi_T(r_1,...)}\delta(r-r_1) [1+\Delta(R)]}_{\psi_T\psi_T}\\
    &= N_e \expval{\frac{\psi_T(r',...)}{\psi_T(r_1,...)}\delta(r-r_1) }_{\psi_T\psi_T} + N_e \expval{\frac{\psi_T(r',...)}{\psi_T(r_1,...)}\delta(r-r_1) \Delta(R)}_{\psi_T\psi_T} \ .
\end{align}
And we get to our final equation by seeing that the first term is just the variational estimate, and the second we call $\gamma^{(1)}_\Delta(r,r')$,
\begin{equation}
    \gamma^{(1)}_M(r,r') =  \gamma^{(1)}_T(r,r') +  \gamma^{(1)}_\Delta(r,r') \ . 
\end{equation}
We can do a similar procedure on our exact estimate, this time we will get two $\Delta$ terms, so we will only take terms linear in $\Delta$.
\begin{align}
    \nonumber     \gamma^{(1)}_0(r,r') &=N_e \expval{\frac{\psi_0(r',...)}{\psi_0(r_1,...)}\delta(r-r_1)}_{\psi_0\psi_0} \\
    \nonumber &=N_e \int dR\ \frac{\psi_0(r',...)}{\psi_0(r_1,...)}\delta(r-r_1)\psi_0(R)\psi_0(R) \\
    \nonumber &=N_e \int dR\ \frac{\psi_T(r',...)[1+\Delta(R')]}{\psi_T(r_1,...)[1+\Delta(R)]}\delta(r-r_1)\psi_T(R)\psi_T(R) [1+\Delta(R)]^2\\
    \nonumber  &=N_e \int dR\ \frac{\psi_T(r',...)}{\psi_T(r_1,...)}\delta(r-r_1)\psi_T(R)\psi_T(R) [1+\Delta(R)] [1+\Delta(R')]\\
    \nonumber  &\approx N_e \expval{\frac{\psi_T(r',...)}{\psi_T(r_1,...)}\delta(r-r_1) [1+\Delta(R)+\Delta(R')]}_{\psi_T\psi_T} \ .
\end{align}
This can again be separated into three terms, the first two are the same as before, and the third one we call $\gamma^{(1)}_{\Delta'}(r,r')$. So up to first order in $\Delta$ our exact estimate is
\begin{equation}
    \gamma^{(1)}_0(r,r') =  \gamma^{(1)}_T(r,r') +  \gamma^{(1)}_\Delta(r,r') + \gamma^{(1)}_{\Delta'}(r,r')  \ . 
\end{equation}
Our FWD estimate will also have second order terms
\begin{equation}
    \gamma^{(1)}_{FWD}(r,r') =N_e \expval{\frac{\psi_T(r',...)}{\psi_T(r_1,...)}\delta(r-r_1) [1+\Delta(R)]^2}_{\psi_T\psi_T} \ .
\end{equation}
And we can see that up to first order in $\Delta$ the FWD estimate is
\begin{equation}
    \gamma^{(1)}_{FWD}(r,r')=  \gamma^{(1)}_T(r,r') +  2\gamma^{(1)}_\Delta(r,r') \ . 
\end{equation}
We can already see that our SUB estimate gives
\begin{equation}
    \gamma^{(1)}_{SUB}(r,r') = 2\gamma^{(1)}_M(r,r')  - \gamma^{(1)}_T(r,r') \ .
\end{equation}
And the DIV estimate is
\begin{equation}
    \gamma^{(1)}_{DIV}(r,r') = \frac{(\gamma^{(1)}_M)^2}{\gamma^{(1)}_T} = \gamma^{(1)}_T + 2\gamma^{(1)}_\Delta + \frac{(\gamma^{(1)}_\Delta)^2}{\gamma^{(1)}_T} \ .
\end{equation}
We can see that for non-local quantities, both SUB and DIV are first order approximations to the FWD result. For a local estimate, like the density, we would have $\Delta(R') \approx \Delta(R)$, so that SUB and DIV would approach the exact solution. This also means that FWD differs from the exact result by the difference between $\Delta(R')$ and $\Delta(R)$.

We can also prove AUX in this manner. Assuming that the wave function can be expressed as Eq.\eqref{equation_corr_estimate_wf}, we can also write,
\begin{equation}
     [1+\Delta(R)] = \frac{\psi_0 (R)}{\psi_T(R) } \propto d(R) \ .
\end{equation}
By using this we can understand our new reweighting scheme as
\begin{align}
    \nonumber \gamma^{(1)}_A(r,r') &=N_e \expval{\frac{\psi_T(r',...)}{\psi_T(r_1,...)}\delta(r-r_1) d(R_i')}_{\psi_T\psi_0}\\
    &= N_e \expval{\frac{\psi_T(r',...)}{\psi_T(r_1,...)}\delta(r-r_1) d(R_i') [1+\Delta(R)]}_{\psi_T\psi_T} \ .
\end{align}
We then use $d(R')\propto [1+\Delta(R')]$ (and since it is a weight it is normalized), we can write the final equation,
\begin{equation}
    \gamma^{(1)}_A(r,r') = N_e \expval{\frac{\psi_T(r',...)}{\psi_T(r_1,...)}\delta(r-r_1)  [1+\Delta(R_i')][1+\Delta(R)]}_{\psi_T\psi_T} \ .
\end{equation}
Which is exactly the same as the ground state 1RDM. It is AUX that introduces the necessary factor $[1+\Delta(R_i')]$, which is the one responsible for the non-local component in the 1RDM.

\section{Lack of cancellation of $\psi_T$ error}
\label{App_psiTcancel}
\begin{figure}[H]
  \centering
  \includegraphics[width=\figwidth]{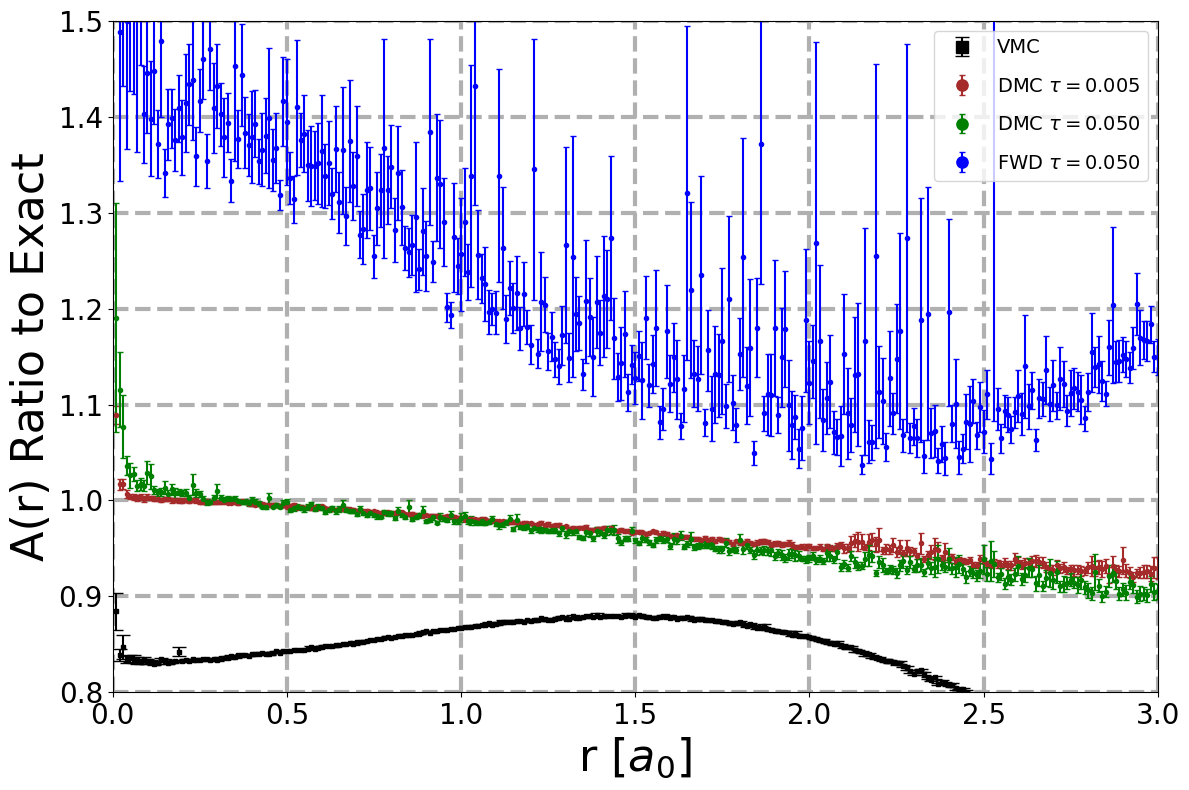}
  \caption{Ratio of the radial part of $A(r)$ computed with QMC to the exact one. A flat line at 1 would represent perfect error cancellation.}
  \label{fig_Ar_benchmark}
\end{figure}

The cancellation of error of the $\psi_T$ in the denominator of the 1RDM accumulator and the distribution is a very fragile process. To the end of studying this cancellation we have decided to compute the following object within the DMC framework
\begin{equation}
    A(r) = 2\expval{\frac{\delta(r-r_1)}{\psi_T(r_1,r_2)}}_{\psi_T\psi_0}\ .
\end{equation}
Which up to a factor $\psi_T(r',r_2)$ is the 1RDM. We have decided to proceed in this way so as to separate the two issues we face when fully unbiasing the 1RDM, the auxiliary reweighting process deals with the non-local part (the numerator), and the cancellation of $\psi_T$ error takes care of the denominator. 
%All our QMC results of $A(r)$ are scaled by the normalization factors shown in equation \ref{eq_Ar_DMC_VMC}. 
This object is spherically symmetric in the Helium atom, so we show the radial part for simplicity. We also had to normalize $\psi_T$ since we only have one, instead of a ratio where the normalization would drop. 
This object is equivalent to
\begin{equation}
    A(r) = 2 \int dr_2 \psi_0 (r,r_2)\ .
\end{equation}
Since in the helium atom we have access to $\psi_0$, we can benchmark our DMC results against the exact solution (For VMC, the overlap integral is 1 as it is just the norm). This is shown in Fig.~\ref{fig_Ar_benchmark}, where we compute the ratio of the QMC $A(r)$ to the exact one. A naive approach to removing the $\psi_T$ error would be to use FWD, as it will lead to sampling with $\psi_0\psi_0$. This will not work, as using FWD will remove the possibility for error cancellation. Using DMC should allow for this cancellation to take place. In Fig.~\ref{fig_Ar_benchmark} can see that DMC improves upon VMC and FWD significantly, but that this cancellation is not realized everywhere. Time step errors are to blame, as their presence stops us from truly sampling $\psi_T\psi_0$, a reduction of time step can be seen to lead to a larger cancellation in Fig.~\ref{fig_Ar_benchmark}. The convergence of the $\psi_T$ cancellation with time step is significantly slower than the energy convergence with time step. The error cancellation is worse at large distances, both because of a time step bias, and because of a lack of walkers in regions where $1/\psi_T$ is increasing in an exponential fashion. These large distances are exactly where AUX or FWD would yield the largest weights.
It is precisely this lack of error cancellation that stops AUX from yielding pure 1RDMs. Still, in the limit of perfect sampling AUX would result in exact 1RDMs, while FWD would not.

\section{Computational details}
\label{App_compdetails}
All calculations shown in this work have been performed with a home made QMC code \cite{rodriguezper_he_atom_qmc}.
Our VMC trial wave function is 
\begin{align}
    \nonumber \psi_T(\mathbf{r}_1,\mathbf{r}_2) = &\exp{-2(\abs{\mathbf{r}_1}+\abs{\mathbf{r}_2})} \\
    &\exp{-f_{ep} \sum_{i}^{N=2}e^{-\abs{\mathbf{r}_i}^2/\omega_{ep}^2} - 
    f_{ee} e^{-r_{12}^2/\omega_{ee}^2} - 
    f_{c} e^{-r_{12}^2/\omega_{c}^2} r_{12}}
\end{align}
Where $f_{ep} = 0.777$, $w_{ep} = 2.56$, $f_{ee} = 0.41$, $w_{ee} = 1.35$, $f_{c} = -0.5$ and $w_{c} = 0.1$ are variational parameters, with their optimized value (The value of $f_{c}$ is fixed by the electron electron cusp). We find a VMC energy of $E_{VMC} =  -2.8911(3)$.

In our DMC simulations of the helium atom we have made use of the importance sampled propagator to evolve our distribution of walkers from the variational $\psi_T^*\psi_T$ to the mixed distribution $\psi_T^*\psi_0$ used to evaluate expectation values at the DMC level. 
\begin{align}
    \nonumber G(\mathbf{R}\leftarrow \mathbf{R}',\tau) &= (2\pi \tau)^{-3N/2} \exp\left(-\frac{[\mathbf{R}-\mathbf{R}'-\tau v_D(\mathbf{R}')]^2}{2\tau}\right) \\
    &\exp \left(-\tau[E(\mathbf{R})+E(\mathbf{R}')-2E_T]/2\right) + \mathcal{O}(\tau^3) \ .
    \label{eq_G_approx}
\end{align}
Each application of this Green's function evolves the distribution by a time step $\tau$, the total diffusion time is $\tau N_{steps}$. The time step introduces an error which can be dealt with by a suitable exptrapolation \cite{umrigar1993diffusion, anderson2024reducing}.
Each configuration $\mathbf{R}$ is called a walker, and they represent the $3N_e$ dimensional coordinates of the $N_e$ electrons in the system. The two terms in Eq.\eqref{eq_G_approx} correspond to two different physical mechanisms. The Gaussian term comes from the kinetic term of the Hamiltonian, with a drift term introduced by the importance sampling. It corresponds to a diffusing move on the walkers $\mathbf{R}$ to a new position $\mathbf{R}'$. The second exponential is introduced by the potential term, and it evaluates the local energy $E_L(\mathbf{R})$ instead of the potential $V(\mathbf{R})$ due to importance sampling. It gives a scalar for each walker which is used to create copies of the walker according to the integer part of the exponential term. This procedure is called branching. 

Forward walking (FWD) makes use of this branching procedure to keep track of the number of descendants of each walker after a sufficient number of steps $N_{FWD}$. The ammount of forward time $\tau_{FWD}=N_{FWD}\cdot \tau$ is kept constant for all our time steps. With the exception of our smallest $\tau=0.005$, where we doubled the forward time to ensure convergence. 

All the QMC calculations presented here have been performed with 40 thousand steps. The number of walkers scales inversely to time step $\tau\cdot N_w = 3000$. The forward time was set to 5 Hartree. VMC was performed with a time step of $0.35$, and 256 thousand walkers. At the larger time steps, FWD was of similar cost to DMC, but the cost of FWD grows faster than DMC as the time step is reduced. Our smallest time step FWD was twice as expensive as the corresponding DMC.

In the Hubbard dimer we took an approach which allowed us to perfectly sample our target distributions. This was possible thanks to the knowledge of $\psi_0$. We could compute the mixed $\psi_T^*\psi_0$ distribution and create walkers distributed exactly according to it. The distribution is just a set of four numbers, so we can multiply them by our target number of walkers and take the integer part. This gives us a set of four integers wihch are the number of walkers in each state, distributed exactly like $\psi_T^*\psi_0$ with a negligible error. A similar procedure is used to sample the variational distribution for VMC calculations. We can also compute analytically the asymptotic number of descendants $d(\mathbf{R})=\psi_0(\mathbf{R})/\psi_T(\mathbf{R})$. For our FWD calculations we used walkers sampled from the mixed distribution which were then reweighted by $d=(\mathbf{R})$. All our VMC, DMC, FWD and AUX calculations of the Hubbard dimer are performed in this way, differing only in the distribution (variational or mixed) and the reweighting scheme (none, FWD or AUX).

\end{appendix}


% Provide your bibliography here. You have two options:

%%% FIRST OPTION
% Write your entries here directly, following the example below, including:
% Author(s), Title, Journal Ref. with year in parentheses at the end, followed by the DOI number.

%\begin{thebibliography}{99}
%\bibitem{1931_Bethe_ZP_71} H. A. Bethe, {\it Zur Theorie der Metalle. i. Eigenwerte und Eigenfunktionen der linearen Atomkette}, Zeit. f{\"u}r Phys. {\bf 71}, 205 (1931), \doi{10.1007\%2FBF01341708}.
%\bibitem{arXiv:1108.2700} P. Ginsparg, {\it It was twenty years ago today... }, \url{http://arxiv.org/abs/1108.2700}.
%\end{thebibliography}

% Use your bibtex library, formatted by the SciPost style file.
%\bibliography{SciPost_Example_BiBTeX_File.bib}

\bibliography{biblio}

%%%%%%%%%% END TODO: BIBLIOGRAPHY

\end{document}